\documentstyle[twoside,fleqn,espcrc2,epsf]{article}

\title{Multiple small angle neutron scattering in ferromagnets}
\author{A.Z.Menshikov, S.G.Bogdanov 
and Y.N.Skryabin\address{Institute for Metal Physics, Russian Academy 
of Sciences, Ural Branch,\\ GSP-170, 620219, Ekaterinburg, Russia}
\thanks{Partially supported by State Program ``Neutron Studies of Matter'',
Russia.}}

\begin{document}

\begin{abstract}
The conclusions of the multiple small angle neutron scattering theory,
developed recently by Maleyev, Pomortsev, and Skryabin (MPS) for the Born
parameter $\alpha \gg 1$ are experimentally confirmed by magnetic scattering
from domain structure of pure Fe and Ni, as well as Fe$_{65}$Ni$_{35}$
alloy.
The crossover
from multiple refraction to multiple Fraunhofer diffraction is found at a
critical thickness, $L_0$, where the neutron beam broadening $w$ vs. sample
thickness $L$ changes from $w\sim \sqrt{L}$ to $w\sim L$.
\end{abstract}

\maketitle

Small angle neutron scattering measurements on unmagnetized ferromagnetic
materials accompany by additional broadening of the incident neutron
beam due to multiple magnetic
scattering on domain structure \cite{hs,wei,shi,bog}.
There are some theoretical
approaches \cite{berk} to describe the  analogical broadening for nonmagnetic
materials in a diffraction range. Recently a new
theory was developed in
\cite{mps} for a refraction regime of scattering, where the mean free path of
neutrons $l$ essentially more than an inhomogeneity radius $R$ ($l\gg R$).

According to this theory the multiple neutron scattering intensity
$I\left ( q,L \right )$ can be considered for two extents of sample thickness
$L \ll L_0$ and $L \gg L_0$, where the critical sample thickness $L_0$ is
given by $L_0 = l\alpha ^2\ln \alpha$
and $\alpha  = Const \lambda R U$.
At $L=L_0$ the crossover from the multiple refraction effect to the multiple
Fraunhofer diffraction turns out. Respectively the broadening of incident
neutron beam
$q_1=\left( \alpha /2R\right) \left[ \left( L/l\right) \ln \left( L/l\right)
\right] ^{1/2}$ is changed to
$q_2=\left( 1/2R\right) \left( 1+1/\pi \right) \left( L/l\right)$.

For magnetic neutron scattering we can write the potential energy as
$U=-\mbox{\boldmath $\mu$}_n \delta \mathbf{B}$,
where $\mbox{\boldmath $\mu$}_n$ is the neutron magnetic moment and $\delta
\mathbf{B}$ $=\mathbf{B}_1-\mathbf{B}_2$ the magnetic contrast.
Considering ferromagnetic domain as inhomogeneity within two component
model, we can write $\mathbf{B}_1=-\mathbf{B}_2$ and
$\delta \mathbf{B}=2\mathbf{B}$, while the mean free path
for spherical inhomogeneities of radius $R_{eff}$ is expressed as
$l=\frac 23R_{eff}\frac V{\delta V}=\frac 43R_{eff}$,
where $V\propto {R_{eff}^3}$ is the sample volume and $\delta V=\frac V2$
the part occupied by inhomogeneities.
Therefore the condition $l>R$ of availability of MPS theory is valid
for a multidomain structure.

To verify the theory \cite{mps} we have performed
experiments on thickness dependence of incident neutron beam broadening
intensities in ferromagnetic pure
iron, nickel and Fe$_{65}$Ni$_{35}$ disordered alloy by using the
small angle neutron
scattering diffractometer with $\lambda =0.5$ nm, installed on IVV-2M reactor.
To obtain the highest broadening of premier neutron beam the plates of samples
were subjected preliminary to a 15-20\% impact-plastic
deformation at room temperature. The magnetic domain structure of
as-deformed iron samples was changes by annealing of samples at 1000 K.
All measured small angle neutron scattering in these materials
have a magnetic nature \cite{bog} because the intensity
disappear when the magnetic saturation state occurs in a sample.

\begin{figure}[htb]
\centerline{\epsfxsize 5.5cm \epsffile{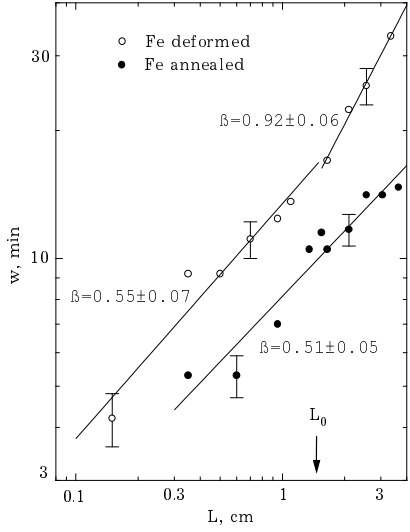}}
\caption{Sample thickness dependence of $w$ for pure Fe.}
\label{fig:1}
\end{figure}

\begin{figure}[htb]
\centerline{\epsfxsize 5.5cm \epsffile{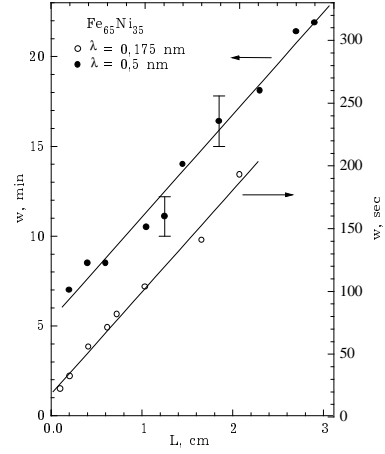}}
\caption{Sample thickness dependence of $w$ for Fe$_{65}$Ni$_{35}$ alloy.}
\label{fig:2}
\end{figure}

Fig.\ref{fig:1} shows neutron beam broadening as a function of thickness for
deformed and annealed iron samples.
Here the broadening is
$w=\frac {\lambda}{2\pi}q= \left( w_1^2-w_0^2\right) ^{1/2}$,
where $w_1$
is the full width at the half-height of the measured beam intensity
maximum, $w_0$ is the incident beam width.
It is seen that the dependence of $w$ on sample
thickness is described by a power law $w\left( L\right) \sim L^\beta $,
which has a break in logarithmic scale for the deformed iron at $L_0=1.55$
cm, where $\beta =1/2$ when $L<L_0$ and $\beta =1$ when $L>L_0$. This break
disappears when we change the size of magnetic domains by annealing.
On the contrary the break occurs for pure Ni
when we change the wave length from $\lambda =0.5$ nm up to
$\lambda =0.16$ nm.
Some standpoint for invar alloy do not change the linear
dependence of the broadening vs. thickness (Fig.\ref{fig:2}).

The incident beam broadening vs. thickness curves
allows evaluation of the effective domain radius and magnetic contrast.
Taking into account the additional relation between $R_{eff}$ and $l$
we have obtained for pure iron $R=18$ $\mu $m
and $\delta B=35$ kGs. It is seen that $\delta B$ occurs to be
close to twice induction $B$ on this materials.
For Fe$_{65}$Ni$_{35}$ alloy
the both linear dependencies $w(L)$ obtained at different wave length of
neutron give us the values of $R_{eff}$ close to each other $R\simeq 19$
$\mu $m ($\lambda =0.5$ nm) and $R\simeq 23$ $\mu $m ($\lambda=0.175$ nm).
Using the experimental induction value $B\simeq 5$ kGs
for this alloy we have obtained
$L_0\simeq 0.06$ cm, a value what is much less than the smallest
sample thickness ($0.2$ cm), used in our measurements.

The authors are thankful to N.Elyutin, S.Matveev and A.Eidlin for the
neutron scattering measurements on double single crystal
spectrometer with $\lambda =0.16$ nm.

\end{document}